\documentclass[twocolumn,showpacs,preprintnumbers,floatfix,prb]{revtex4}

\usepackage{amsmath,amssymb}
\usepackage{xspace}
\usepackage{graphicx}
\usepackage{dcolumn}
\usepackage{bm}
\input{MyMc.tex}
\newcommand{\ZeroPi}{{\ensuremath{0\text{-}\pi}}\xspace}

\begin{document}

\title{%
  Escape Rate Measurements and Microwave Spectroscopy of\\
  $0$, $\pi$, and \ZeroPi ferromagnetic Josephson Tunnel Junctions
}

\author{J. Pfeiffer}
\email{judith.pfeiffer@uni-tuebingen.de}
\author{T. Gaber}
\author{D. Koelle}
\author{R. Kleiner}
\author{E. Goldobin}
\affiliation{%
  Physikalisches Institut-Experimentalphysik II and Center for Collective Quantum Phenomena, 
  Universit\"at T\"ubingen, D-72076 T\"ubingen, Germany
}

\author{M. Weides}
\altaffiliation{Current adress: Department of Physics, University of California, Santa Barbara, CA 93106, USA}
\author{H. Kohlstedt}
\affiliation{%
  Institute of Solid State Research and JARA-Fundamentals
  of Future Information Technology, Research Centre, J\"ulich,
  D-52425 J\"ulich, Germany
}

\author{J.~Lisenfeld}
\author{A.~K.~Feofanov}
\author{A. V. Ustinov} 
\affiliation{%
  Physikalisches Institut, Universit\"at Karlsruhe, D-76131 Karlsruhe, Germany
}

\date{\today}

\begin{abstract}

  We present experimental studies of high quality underdamped $0$, $\pi$, and \ZeroPi ferromagnetic Josephson tunnel junctions of intermediate length $L$ ($\lambda_J \lesssim L \lesssim 5\lambda_J$, where $\lambda_J$ is the Josephson penetration depth). The junctions are fabricated as Nb/Al$_2$O$_3$/Cu$_{40}$Ni$_{60}$/Nb Superconductor-Insulator-Ferromagnet-Superconductor heterostructures. Using microwave spectroscopy, we have investigated the eigenfrequencies of $0$, $\pi$, and \ZeroPi Josephson junctions in the temperature range 1.9\,K...320\,mK.  Harmonic, subharmonic and superharmonic pumping is observed in experiment, and the experimental data are compared with numerical simulations. Escape rate measurements without applied microwaves at temperatures $T$ down to 20\,mK show that the width of the switching current histogram decreases with temperature and saturates below $T=150\units{mK}$. We analyze our data in the framework of the short junction model. The differences between experimental data and theoretical predictions are discussed.
 
\end{abstract} 



\maketitle

\section{Introduction}
\label{Sec:Intro}

Josephson junctions with a phase drop of $\pi$ in the ground state\cite{Bulaevskii:pi-loop}, so called $\pi$ junctions, are intensively investigated, as they promise important advantages for Josephson junction based electronics\cite{Terzioglu:1997:CompJosLogic,Terzioglu:1998:CJJ-Logic}, and, in particular, for Josephson junction based qubits\cite{Ioffe:1999:sds-waveQubit,Blatter:2001:QubitDesign,Yamashita:2005:pi-qubit:SFS+SIS,Yamashita:2006:pi-qubit:3JJ}. Nowadays, several technologies allow to manufacture such junctions: Josephson junctions with a ferromagnetic barrier\cite{Ryazanov:2001:SFS-PiJJ,Oboznov:2006:SFS-Ic(dF),Kontos:2002:SIFS-PiJJ,Weides:2006:SIFS-HiJcPiJJ}, quantum dot junctions\cite{vanDam:2006:QuDot:SuperCurrRev,Cleuziou:2006:CNT-SQUID,Jorgensen:2008:QuDotJJ:0-pi-transition} and nonequilibrium superconductor - normal metal - superconductor Josephson junctions\cite{Baselmans:1999:SNS-pi-JJ,Huang:2002:NonEquPiJJ}.

Furthermore, one can fabricate 0-$\pi$ long Josephson junctions \cite{Tsuei:Review,Kirtley:SF:HTSGB,Lombardi:2002:dWaveGB,Smilde:ZigzagPRL,Weides:2006:SIFS-0-pi}, \ie, junctions with some parts behaving as 0 junctions and other parts behaving as $\pi$ junctions. The ground state phase $\mu(x)$ in such junctions has a value of 0 deep inside the 0-region, and a value of $\pi$ deep inside the $\pi$ region. At the 0-$\pi$ boundary it continuously changes from 0 to $\pi$ on the scale of $\lambda_J$, where $\lambda_J$ is the Josephson penetration depth. Such a bending of the phase results in the appearance of a local magnetic field $\propto d\mu/dx$. As supercurrents $\pm\sin[\mu(x)]$ circulate around the boundary, one deals with a pinned Josephson vortex. Its total magnetic flux $\Phi$ is equal to $\Phi_0/2$ for $L\gg\lambda_J$, where $L$ is the length of the junction and $\Phi_0\approx2.07\times10^{-15}\units{Wb}$ is the magnetic flux quantum. Such a Josephson vortex is called a semifluxon\cite{Bulaevskii:0-pi-LJJ,Goldobin:SF-Shape,Xu:SF-Shape}. If the Josephson phase $\mu(x)$ deep inside the $\pi$ region is equal to $-\pi$ instead of $\pi$, the localized magnetic flux is equal to $-\Phi_0/2$ and the supercurrent of the vortex circulates counterclockwise (antisemifluxon). Both semifluxons and antisemifluxons were observed experimentally\cite{Hilgenkamp:zigzag:SF,Kirtley:2005:AFM-SF} and have been under extensive experimental and theoretical investigation during the last decade\cite{Kogan:3CrystalVortices,Kirtley:SF:HTSGB,Kirtley:SF:T-dep,Hilgenkamp:zigzag:SF,Kirtley:IcH-PiLJJ,Goldobin:SF-ReArrange,Stefanakis:ZFS/2,Zenchuk:2003:AnalXover,Goldobin:Art-0-pi,Susanto:SF-gamma_c,Goldobin:2KappaGroundStates,Goldobin:F-SF,Buckenmaier:2007:ExpEigenFreq,Nappi:2007:0-pi:Fiske}. If the 0-$\pi$ Josephson junction length $L\lesssim\lambda_J$, the semifluxon does not fully fit into the junction and the flux $|\Phi|<\Phi_0/2$. 

Superconductor-Insulator-Ferromagnet-Super\-con\-duc\-tor (SIFS) Josephson tunnel junctions are interesting devices for quantum applications\cite{Ioffe:1999:sds-waveQubit,Yamashita:2005:pi-qubit:SFS+SIS} because the $\pi$ phase shift is provided ``for free'' without the need of extra gate electrodes\cite{vanDam:2006:QuDot:SuperCurrRev,Cleuziou:2006:CNT-SQUID,Jorgensen:2008:QuDotJJ:0-pi-transition} or current injectors\cite{Baselmans:1999:SNS-pi-JJ,Huang:2002:NonEquPiJJ,Goldobin:Art-0-pi}. In addition, unlike for $d$-wave\cite{Kirtley:SF:HTSGB,Smilde:ZigzagPRL} or SFS\cite{DellaRocca:2005:0-pi-SFS:SF,Frolov:2006:SFS-0-pi} based 0-$\pi$ Josephson junctions, the dissipation in SIFS junctions decreases exponentially at low temperatures\cite{Weides:2006:SIFS-HiJcPiJJ,Pfeiffer:2008:SIFS-0-pi:HIZFS} as in conventional Josephson tunnel junctions. In this paper, we study phase fluctuations and dynamics in a SIFS \ZeroPi Josephson junction and compare it with its two reference $0$ and $\pi$ junctions, at temperatures between 1.9\,K down and 20\,mK. The different escape mechanisms of the phase --- by thermal activation (TA) and resonant activation (RA) --- are studied experimentally for junctions of intermediate length ($\lambda_J \lesssim L \lesssim 5\lambda_J$).

\section{Samples and Measurement Techniques}
\label{Sec:ExpRes}

\begin{figure*}[!tb]
  \begin{center}
    \includegraphics[width=15cm]{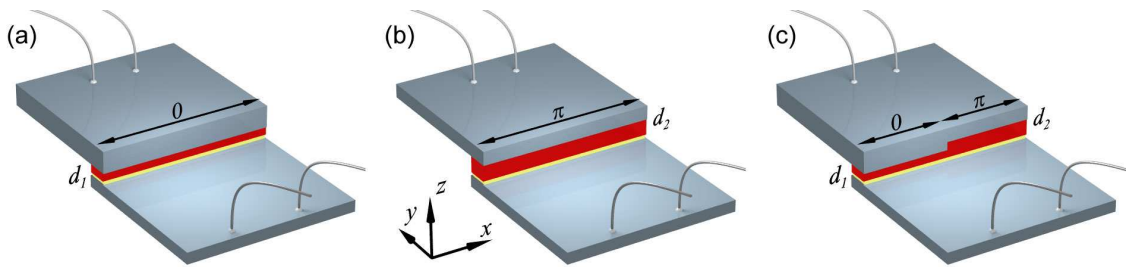}
 \end{center}
  \caption{(Color online) 
    Schematic drawing of $0$ (a), $\pi$ (b) and \ZeroPi (c) SIFS Josephson junctions in overlap geometry. Top and bottom electrodes are coloured in light blue (gray), the insulating barrier is yellow (light gray), the ferromagnetic barrier is red (dark gray). For the reference $0$ junction $d_F=d_1$ (a), for the $\pi$ junction $d_F=d_2$ (b). The ferromagnetic barrier of the \ZeroPi junction has a step-like change in the thickness $d_F$ along the $x$ axis (c). 
      }
  \label{Fig:Sketch}
\end{figure*}


\begin{table}[b]
  \centering
  \begin{tabular}{l|c c c c}
    \hline\hline    
    \textit{id} @ $T$ (K) 
    & \ensuremath{j_c} ($\units{A/cm^2}$) 
    & \ensuremath{\lambda_J} ($\units{\mu m}$)
    & \ensuremath{l} \\

    \hline
    $0$ @ 0.32 K    &13.4 &160  &3.1 \\
    $\pi$ @ 0.32 K    &4.5  &280  &1.8\\
    \ZeroPi @ 0.32 K  &9.0  &200  &2.5 \\
    \hline\hline    
  \end{tabular}
  \caption{%
    Parameters of the samples. $j_c$ is the critical current density, $\lambda_J$ is the Josephson penetration depth, $l=L/\lambda_J$ is the normalized length. 
  }
  \label{Tab:Params}
\end{table}

The SIFS Josephson junctions studied are fabricated in overlap geometry using Nb/Al-Al$_2$O$_3$/Ni$_{60}$Cu$_{40}$/Nb technology\cite{Weides:2007:JJ:TaylorBarrier,Weides:SIFS-Tech,Weides:2006:SIFS-HiJcPiJJ,Weides:2006:SIFS-0-pi}, see Fig.~{\ref{Fig:Sketch}}. Depending on the thickness $d_F$ of the ferromagnetic barrier, the junctions can be either in a $0$ or a $\pi$ ground state (usually $d_F$ is of the order of several nm, \eg 3\dots 8\,nm)\cite{Weides:2006:SIFS-HiJcPiJJ}. For reference, the chip contains a $0$ junction with $d_F=d_1$, see Fig.~{\ref{Fig:Sketch}} (a), and a $\pi$ junction with $d_F=d_2$, see Fig.~{\ref{Fig:Sketch}} (b). To fabricate \ZeroPi SIFS Josephson junctions, the ferromagnetic layer was selectively etched along one half of the junction. In this way one half of the junction has an F-layer thickness $d_F=d_1$ having the ground state $\mu=0$ (if taken seperately), while the other half of the junction has an F-layer thickness $d_F=d_2$ having a ground state with $\mu=\pi$, see Fig.~{\ref{Fig:Sketch}} (c)\cite{Weides:2006:SIFS-0-pi}. The lengths of the $0$ and $\pi$ parts are equal within the limit of lithographic accuracy of $\sim 1\units{\mu m}$. The physical length of each of the three junctions is $L=500\,\mu$m, the width is $W=12.5\,\mu$m. The parameters of the samples are summarized in Tab.~\ref{Tab:Params}. The critical current densities of the reference junctions are obtained by measuring their critical current dependencies on magnetic field, $I_c(B)$. For the \ZeroPi Josephson junction the value $j_c^{\ZeroPi}=(|j^{\pi}_c| + j^0_c)/2$ is quoted and its normalized length $l=L/\lambda_J$ is calculated from this value. Here, we assumed that the critical current densities in the $0$ and $\pi$ parts are the same as for the respective reference junctions located nearby. While calculating $l$ the idle region corrections are taken into account.\cite{Wallraff:PhD,Monaco:1995:IdleReg:Dyn} Detailed information on the dynamic and static properties of these samples has been published elsewhere\cite{Pfeiffer:2008:SIFS-0-pi:HIZFS}. For temperatures $T>3.5$\,K the junctions were overdamped, exhibiting non-hysteretic current-voltage characteristics. At lower temperatures they became increasingly underdamped. The measurements we present here were done in the underdamped regime.


The measurements are carried out in (a) a standard $^3$He cryostat and (b) a dilution refrigerator. Using the $^3$He cryostat, temperatures between $1.9\units{K}$ and $320\units{mK}$ are accessible, while the dilution refrigerator has a base temperature of about 20\,mK. In both setups, cryoperm shields are placed around the sample to shield it from the earth magnetic field and stray fields. In the $^3$He setup an external magnetic field can be applied in-plane of the junctions in a controlled way by using a solenoid. In the dilution refrigerator a bonding wire is used to expose the junction to magnetic field. As the samples are very sensitive to residual magnetic fields, special care has been taken to ensure a parasitic-flux-free state. A series of filter stages is used for both the bias current and the voltage sensing lines. At different temperature stages commercial feed-through filters, RC filters and custom-made capacitively shunted copper powder filters are used\cite{Liesenfeld:PhDThesis,PowderFilters}. A current divider of the ratio 1:56 is inserted in the current line to improve the signal-to-noise ratio.

We investigate the escape of the Josephson phase by measuring switching current statistics with a time-domain technique\cite{Fulton:SignaPropToIcToOneThird}. In brief, the sample is current biased by a custom-made low-noise battery-powered current source. The bias current is ramped up starting from $I=0$ at a time $t=0$ at a constant rate $\dot{I}$. At the time $t_{\rm sw}$, which is measured using a counter with a 20\,GHz stabilized clock, the voltage detector detects that the junction switches to a finite voltage state. Each switching event is detected by feeding the preamplified voltage signal from the sample to a custom-made trigger circuit with adjustable threshold. The switching current $I_\mathrm{sw}$ is calculated as $I_\mathrm{sw}=\dot{I}\times t_{\rm sw}$. The switching current probability distribution $P(I_\mathrm{sw})$ is found by accumulating a large number ($\sim 10^4$) of measurements of $I_\mathrm{sw}$ and generating a histogram. The standard deviation $\sigma$ of $P(I_\mathrm{sw})$ is further evaluated as a function of temperature. 

To determine the eigenfrequencies of the junctions spectroscopy measurements are performed. For this purpose, microwaves in the frequency range between 1 and 15\,GHz are applied to the junctions by placing an antenna above the sample. The microwave power (at the output of the source) was varied in the range of $-$100...20\,dBm. A detailed description of the setups and used measurement techniques is published elsewhere\cite{Liesenfeld:PhDThesis,walli_setup,walli_PRL90}.

\section{Escape Rate Measurements}
\label{subSec:EscMea}

\begin{figure}[tb]
  \includegraphics[width=4cm]{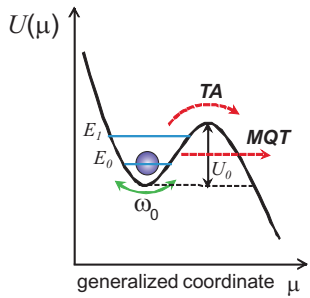}
  \caption{(Color online)
    Virtual particle of mass $m$ in a tilted washboard (cosine) potential $U(\mu)$. $\omega_{0}$ is the small amplitude oscillation frequency (eigenfrequency), $U_0$ the barrier height. The energy levels in the well are indicated, likewise the two escape mechanisms (TA: thermal activation, MQT: macroscopic quantum tunnelling).
  }
  \label{schema_TA_MQT}
\end{figure}

The phase dynamic of \emph{small} $0$ Josephson junctions ($L<\lambda_J$) was extensively studied in the literature\cite{Likharev:JJ&Circuits}. It is described by the resistively and capacitively shunted junction (RSCJ) model. The same equations describe the dynamics of a point-like particle of mass $m$ moving along the coordinate $\mu$ in a tilted wash-board potential $U(\mu)=(1-\cos\mu)-\gamma\mu$, see Fig.~{\ref{schema_TA_MQT}}. The potential is tilted by the applied (normalized) bias current $\gamma$ and has local minima for $|\gamma|<1$. If the particle is trapped in one of the minima (zero voltage state), it may perform small amplitude oscillations around the bottom of the well at the eigenfrequency 
\begin{equation}
  \omega_0(\gamma)=\omega_p(1-\gamma^2)^{1/4}
  \label{omega0}
\end{equation}
with $\omega_p=(2\pi I_c/\Phi_0 C)^{1/2}$ being the Josephson plasma frequency. $I_c$ is the critical current of the junction, $C$ its capacitance. When $|\gamma|\to 1$ the minima disappear (the energy barrier vanishes) and the particle starts moving with non-zero average velocity $\dot{\mu}$ (non-zero voltage). In the presence of thermal or quantum fluctuations the particle may escape from the potential well even for $|\gamma|\lesssim 1$. In the \emph{thermal regime}, the particle is thermally activated over the potential barrier\cite{Haengii}. The probability of such a process follows the Boltzmann distribution and thus strongly depends on temperature. At low temperatures the thermal escape probability becomes very small, so that macroscopic \emph{quantum tunnelling} (MQT) out of the well becomes dominant. Yet another escape mechanism is to resonantly excite the particle by an external ac force at a frequency close to its eigenfrequency (Eq.~(\ref{omega0})), as it is done in microwave spectroscopy\cite{Devoret}. When the particle escapes from the potential well, provided the damping is not too high, it slides down the washboard potential and does not stop in the next wells resulting in a finite voltage state.

If one considers similar processes in a \emph{long} Josephson junction, the spatial dependence of the Josephson phase $\mu(x)$ has to be taken into account. A long junction is not described by a single particle moving in a washboard potential, but by a string. This string can overcome the barrier as a whole, having the same phase not dependent on the coordinate $x$ at each moment, but it can also bend, so that first one part of the string passes over the barrier and then it pulls the rest\cite{Castellano:1996:LJJ:ThermEsc}. Such activation processes for linear and annular long Josephson junctions were studied theoretically\cite{Castellano:1996:LJJ:ThermEsc,Fistul:Qu-F-AF,Gulevich:2006:SwitchAJJ} and experimentally\cite{Castellano:1996:LJJ:ThermEsc,Fistul:Qu-F-AF}. For a infinitely long \ZeroPi Josephson junction containing a fractional vortex, the theory of thermal activation and MQT was developed recently\cite{vogel-2008}. MQT of a fluxon was described\cite{Kato:1996:MQT-FluxonLJJ} and already observed\cite{Wallraff:2003:Fluxon:QuTu}.

\begin{figure}[tb]
  \includegraphics[width=7.8cm]{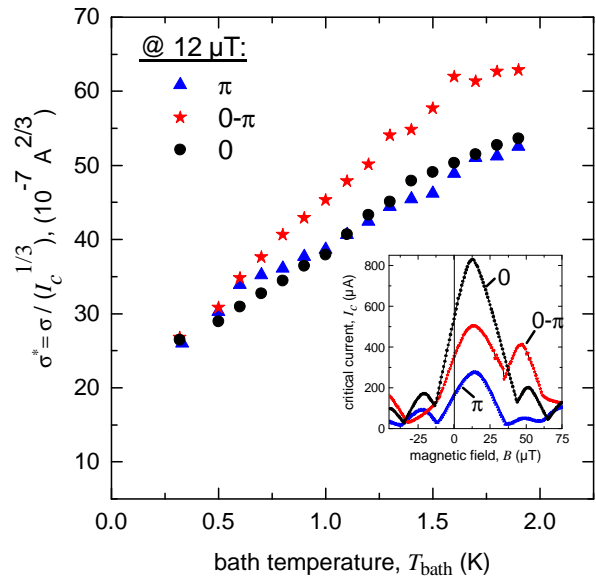}
  \caption{(Color online) 
    Experimental data of phase escape rate measurements of the 0, $\pi$ and \ZeroPi Josephson junction at 12\,$\mu$T: Standard deviation $\sigma^*$  between $320\units{mK}\ldots1.9\units{K}$ is plotted in reduced units. Inset: Critical current vs. applied magnetic field of the three samples at $T=320\,$mK. 
}
  \label{sigma(T)_alle_drei_He3}
\end{figure}

Fig.~{\ref{sigma(T)_alle_drei_He3}} shows data for our three samples, measured in the $^3$He cryostat. To characterize the samples, $I_c(B)$ measurements were taken at $T=320\,$mK, see inset of Fig.~{\ref{sigma(T)_alle_drei_He3}}. Both reference junctions exhibit an almost regular Fraunhofer pattern indicating no substantial amount of parasitic flux. Note however, that both curves are not centered around zero magnetic field, but shifted by $\sim12\,\mu$T. This shift and also the slight asymmetries in the side lobes of the Fraunhofer pattern occur due to the magnetization of the F-layer\footnote{M. Kemmler {\it et al.}, unpublished, 2009.} and were different in every cooldown. For the \ZeroPi junction the characteristic dip in the middle of the $I_c(B)$ is visible, being a signature of the \ZeroPi boundary\cite{Weides:2006:SIFS-0-pi}. By conincidence, the left maximum at $\sim12\,\mu$T coincides with the maxima of the two reference junctions.

\begin{figure}[tb]
  \includegraphics[width=7.8cm]{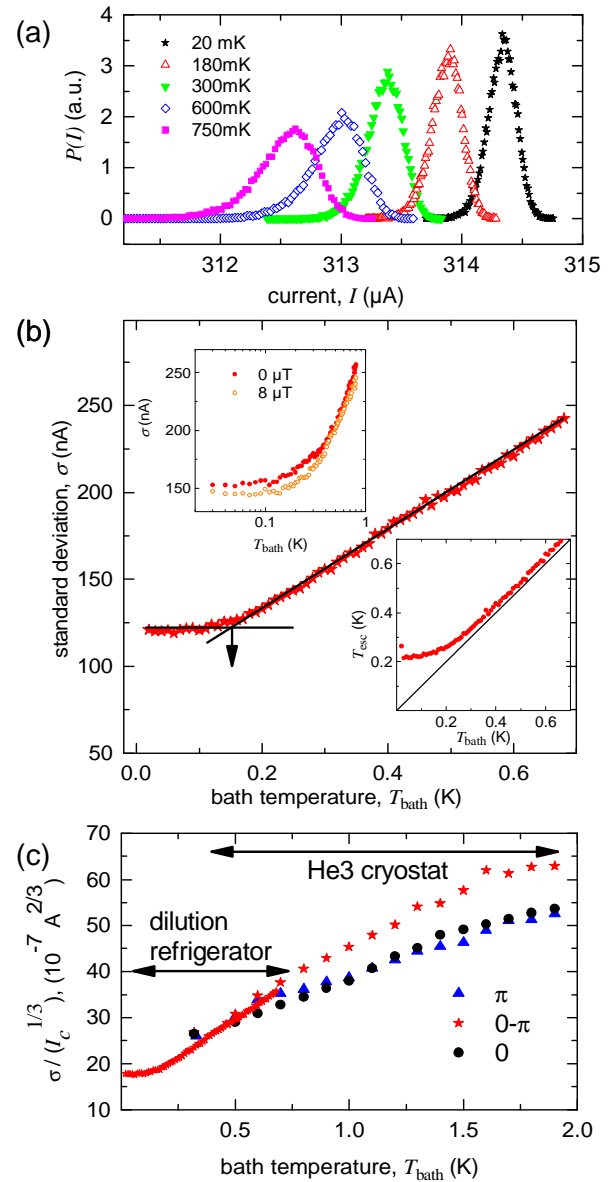}
  \caption{(Color online) 
    Escape rate measurements of the \ZeroPi junction for temperatures between $20\units{mK}$ and $\sim 1\units{K}$. (a) Switching current distributions $P(I)$ for five different temperatures. (b) Standard deviation $\sigma$ of $P(I)$ versus $T$ showing a saturation of $\sigma$ below 150\,mK (measuring cycle \#1). Lower inset: $T_{{\rm esc}}$ vs. $T_{{\rm bath}}$ of the same experimental data (measuring cycle \#1). The bath temperature is indicated as a black line. Upper inset: $\sigma$ vs. $T$ with and without magnetic field applied (measuring cycle \#2). (c) Comparison of the different setups: measurements in the $^3$He cryostat and the dilution refrigerator are plotted in reduced units.
}
  \label{crossover_m12}
\end{figure}

The escape rate measurements are performed at these maxima of the $I_c(B)$ patterns, in order to minimize $I_c$ fluctuations due to fluctuations of the applied magnetic field. Note that comparable measurements are difficult in the minimum of $I_c(B)$ of the \ZeroPi junction as the minimum position slightly shifts with temperature. Therefore, the bias point in field should be adjusted for each temperature value. This behaviour was also observed in other long \ZeroPi junctions. The main panel of Fig.~{\ref{sigma(T)_alle_drei_He3}} shows, as the central result of this section, the standard deviation $\sigma$ vs. $T$ for all three junctions. The probability distribution $P(I)$ of the switching currents has been obtained by recording $2 \times 10^4$ individual switching events at each temperature. The current was ramped with a rate of $\dot{I}\sim 0.4$\,A/s. As all three junctions have different critical currents a direct comparison of $\sigma$ is not possible. However, in theory\cite{Fulton:SignaPropToIcToOneThird,Jackel,Garg} a scaling of $\sigma\sim I_c^{1/3}$ is expected, thus for sake of comparison $\sigma^*(T)=\sigma/I_c^{1/3}$ is plotted in reduced units. The histogram width becomes more narrow when the temperature is decreased, see Fig.~{\ref{sigma(T)_alle_drei_He3}} and compare with Fig.~{\ref{crossover_m12}} (a), as the thermal fluctuations decline. Down to 320\,mK the standard deviation decreases with temperature as it is expected in the thermal regime. 

The $0$ and $\pi$ reference junctions show the same reduced $\sigma^*$ values, \ie, the noise in SIFS Josephson junctions does not depend on $d_F$. The $\sigma^*$ values of the \ZeroPi junction are higher than the ones of the reference junctions in the temperature range between 1.9\,K...500\,mK. This additional noise might be caused by the fluctuations of the fractional vortex located at the \ZeroPi boundary, as also observed for Nb/Al-Al$_2$O$_3$/Nb injector junctions\footnote{Uta Kienzle {\it et al.}, unpublished, 2009.}.

Fig.~{\ref{crossover_m12}} (main panels) shows escape rate data of the \ZeroPi junction measured down to 20\,mK in a dilution refrigerator in the absence of applied magnetic field. The current was ramped at a rate of $\dot{I}\sim 22$\,A/s. In this measurement the critical current of the junction was $\sim 40$\% lower than its maximum value, cf. inset of Fig.~{\ref{sigma(T)_alle_drei_He3}}, \ie, the bias point in field is different from the one in Fig.~{\ref{sigma(T)_alle_drei_He3}}. Fig.~{\ref{crossover_m12}}  (a) shows representative histograms at five different temperatures. In Fig.~{\ref{crossover_m12}} (b) $\sigma$ vs. $T$ is plotted for a large number of temperature values. For $T \ge 150\units{mK}$ the junction is in the thermal regime, where $\sigma$ decreases with decreasing temperature. For $T \le 150\units{mK}$ the standard deviation saturates and stays at a constant level of $\sim 125$\,nA. In order to estimate whether this saturation indicates a crossover to the quantum regime or is due to other reasons, we roughly estimate the crossover temperature using the short junction expression\cite{MDC87:ShortJJ_Expr}

\begin{equation}
  T^*_{{\rm theo}}\sim\frac{\hbar\omega_{0}(\gamma^*)}{2\pi k_B}.
\label{crossover}
\end{equation}
$k_B$ is the Boltzmann constant and $\omega_{0}(\gamma^*)$ is the eigenfrequency, evaluated at the most probable switching current $\gamma^*$, obtained by fitting the histograms according to Ref.~[\onlinecite{walli_setup}]. Using the experimentally determined parameters for the \ZeroPi junction we get $T^*_{{\rm theo}}\sim 6\dots 10\units{mK}$ depending on $\gamma^*$. This value is almost fifteen times smaller than the experimentally observed crossover temperature of $T^*_{{\rm exp}}\sim 150\units{mK}$. Actually, $T^*_{{\rm theo}}\sim 10\units{mK}$ would not be reachable in our measurement setup.
\\

Due to this large discrepancy one might be inclined to suspect current noise in our setup as being the limiting factor. Below we present several arguments against this assumption.
\\

(a) We analyzed the experimental data by fitting the switching current distribution, leaving the temperature as a free parameter according to Ref.~[\onlinecite{walli_setup}]. $T_{{\rm esc}}$ is the effective temperature which is calculated to be consistent with the observed escape rate dependence on bias current. First, as in the classical transition state theory no damping is considered in the data analysis. The results are shown in the lower inset of Fig.~{\ref{crossover_m12}}(b). The calculated escape temperatures are close to the bath temperature. In the whole temperature range discussed, $T_{{\rm esc}}$ is only $\sim30\dots 50\,$mK higher than $T_{{\rm bath}}$, indicating that the measurements are not substantially affected by enviromental noise. Repeating the data analysis assuming intermediate and moderate-to-high damping\cite{kempi} and thus modifying the escape rate by a prefactor yields $T_{{\rm esc}}$ which is different from $T_{{\rm bath}}$ by the same amount of 30$\dots$50\,mK. 

(b) To rule out electronic noise as the limiting factor we additionally applied magnetic field to suppress $I_c$ and thus reduce $\sigma$. The upper inset of Fig.~{\ref{crossover_m12}}(b) shows two series of escape rate measurements which were taken in the second measurement cycle (\#2). Though the saturation level here is different, they show a similar crossover temperature $T^*_{{\rm exp}}\sim 150$\,mK. The measurements were performed without (filled circles) and with (open circles) applied magnetic field. In accordance with the measured $I_c$ reduction of 6\%, $\sigma$ decreases by 2$\dots$3\% as theoretically expected. We further note that the $\sigma$ values for these two measurements are higher than in the main panel of Fig.~{\ref{crossover_m12}}(b). However, the reduced standard deviation $\sigma^*$ was the same for all different measurement cycles. Thus, by applying magnetic field the standard deviation changes in agreement with expectations, again indicating that our measurements are not limited by electronic noise. 

(c) In the dilution refrigerator setup used a current noise level as low as 20$\dots$30\,nA\cite{walli_nature} was measured. Thus, our observed saturation level of $\sim 125$\,nA is not the resolution limit of this setup.

(d) $\sigma^*(T)$ obtained in the $^3$He cryostat coincides with the one measured in the dilution refrigerator, see Fig.~{\ref{crossover_m12}}(c), in the temperature range between 700...300\,mK which is available in both setups. The data taken in the dilution refrigerator smoothly merge to the $^3$He cryostat data, there is no systematic shift between the data sets. 

(e) Also, for the $0$ reference junction we obtained an unexpected high crossover temperature $T^*_{{\rm exp}}\sim 110\ldots120\units{mK}$ (not shown). Unfortunately we could not measure the $\pi$ junction in the dilution refrigerator due to sample problems in several measurement cycles. However, there are data of a $\pi$ junction obtained by K.~Madek \textit{et al.}\footnote{K. Madek {\it et al.}, unpublished. R. Gross, private communication.} on samples produced with the same technology (using the same machine). The authors found $T^*_{{\rm exp}}\sim$100\,mK, which is in accord with our results for 0 and 0-$\pi$ SIFS Josephson junctions. 

Overall, we conclude that the observed saturation in the standard deviation of the switching current distributions is not caused by the measurement setup, but is an intrinsic property of our SIFS Josephson junctions. Additional studies have to address the question whether the type of ferromagnet or the thickness of the ferromagnetic layer has an influence on the results, \ie, whether and how the unexpected high crossover temperature depends on the ferromagnetic layer. Furthermore, a theoretical model is needed to describe the crossover to the quantum regime in SIFS structures. First, the short junction SIS model is an appropriate model for rough estimations and for not very long Josephson junctions. Second, one can take into account the effect of the fluctuations in the ferromagnet on the Josephson phase dynamics. In addition, samples of different normalized length ($l\ll 1$ and $l\gg 1$) are also of high interest, either to avoid length effects or to concentrate on the dominant fluctuations at the \ZeroPi boundary.

\section{Microwave Spectroscopy}
\label{subSec:MicSpec}

\begin{figure}[!tb]
  \includegraphics[width=6.6cm]{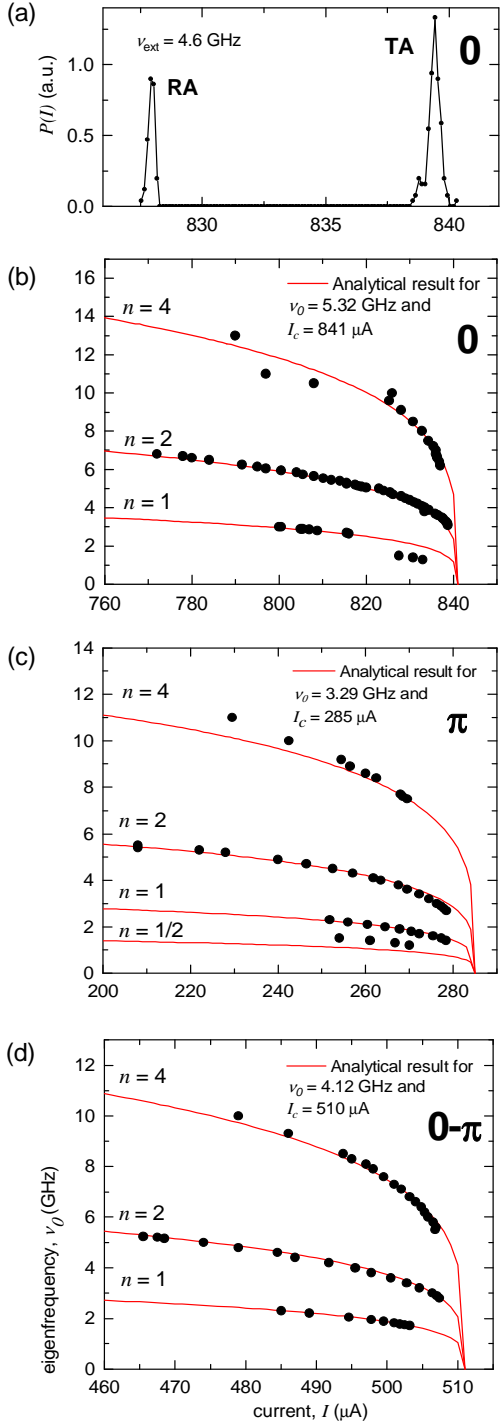}
  \caption{(Color online) (a) Example of a switching current histogram with two peaks, the resonant activation peak (RA) and the thermal activation peak (TA). The microwave frequency is $\nu_{\rm ext}=4.6$\,GHz at $T=320$\,mK. $\nu_0(\gamma)$ dependencies for $0$ (b), $\pi$ (c) and \ZeroPi (d) Josephson junction. Harmonic ($n=1$), subharmonic ($n=1/2$) and superharmonic ($n=2,\,4$) branches are indicated. The parameters $I_c$ and $\nu_0$ of the analytical fits (solid red lines) are given. Black dots show experimental data.
}
  \label{spectroscopy}
\end{figure}

\begin{figure}[!tb]
  \includegraphics[width=7cm]{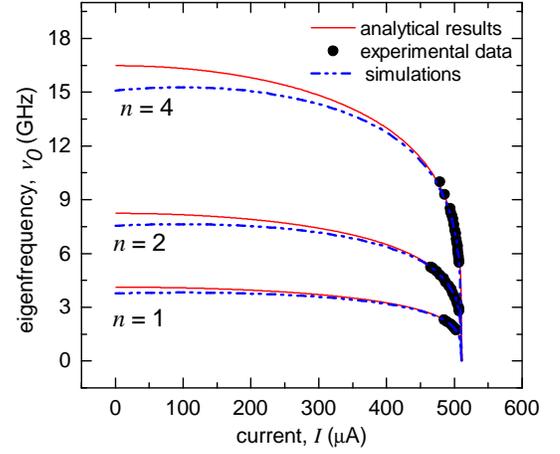}
  \caption{(Color online) 
    Comparison between experimental data (circles), analytical fit (red, solid line) and numerical simulations (blue, dash dotted line) of the microwave spectroscopy data of the \ZeroPi junction at 320\,mK. 
  }
  \label{spectroscopy_vergl_simu}
\end{figure}

Additional studies on the phase dynamics of the 0, $\pi$ and \ZeroPi Josephson junctions were performed by microwave spectroscopy. Here, the phase is resonantly excited from the potential well by an external ac force. The experimental data were obtained by irradiating the junctions with microwaves of frequencies close to $\nu_0=\omega_0/2\pi$, \cf Eq.~({\ref{omega0}}). As a starting point $\omega_p$ was estimated with the capacitance $C$ obtained from Fiske step measurements and $I_c$ from the respective $I_c(B)$ pattern. Switching current measurements were carried out simultaneously. The measurements were performed for different microwave power levels. At low power values the switching current distribution is comparable to the one without microwaves --- no resonant excitation is observed. At some higher power values (the particular value depends on the coupling between the antenna and the sample) in addition to the thermal escape peak around $I_c$ a resonant activation peak becomes visible at lower switching currents, see Fig.~{\ref{spectroscopy}} (a). By tracking the current position of the resonant peak for different frequencies of the external drive $\nu_{\rm ext}$, the eigenfrequencies $\omega_0(\gamma)$ of the samples can be obtained. This is done under the condition that the amplitudes of the resonant peak and of the initial escape peak are more or less equal.

As for the escape rate measurements  (\cf Fig.~{\ref{sigma(T)_alle_drei_He3}} ) the junctions were flux biased at their maxima of the $I_c(B)$ patterns (in fact, both, escape rate and microwave spectroscopy measurements were done in the same run). The microwave spectroscopy is done at $T=320\,$mK, in the thermal regime. 

Fig.~{\ref{spectroscopy}} shows the resulting eigenfrequencies as a function of bias current. We observed harmonic pumping ($\sim\nu_0$), superharmonic pumping ($\sim n\times \nu_0$, $n$ being an integer number) and subharmonic pumping ($\sim \frac{\nu_0}{n}$). To analyze the spectroscopy data, we used for fitting the eigenfrequencies according to the short junction model 
\begin{equation}
\nu_{n}=\frac{n\omega_0(\gamma)}{2\pi}.
\label{nu}
\end{equation}
$I_c$ and $\nu_0$ are taken as fitting parameters. The results are shown in Fig.~{\ref{spectroscopy}} as solid red lines. The respective values for $I_c$ and $\nu_0$ are indicated. Surprisingly, although the samples are in an intermediate length limit, and are SIFS rather than SIS samples, the fit according to the short junction model (Eqs.~(\ref{omega0}) and (\ref{nu})) reproduces the experimental data accurately. Especially for the \ZeroPi junction one could have expected a signature of the fractional flux pinned at the 0-$\pi$  boundary, \cf Fig.~{\ref{sigma(T)_alle_drei_He3}}, resulting in a discrepancy between the eigenfrequencies according to the short junction model and the experimental data. 

For further analysis, we performed numerical simulations of the eigenfrequencies of the \ZeroPi junction. These simulations take the different $j_c^0\neq j_c^{\pi}$ in the $0$ and $\pi$ regimes and the finite length of the junction into account. For given junction parameters and fixed bias current, in a first step we find a stable static\footnote{In fact we solve the dynamic sine-Gordon equation and wait until the solution has relaxed to a static one.} solution $\mu_0(x)$ of the sine-Gordon equation 
\begin{equation}
  \mu_{xx} - j_c(x)\sin(\mu) = \gamma
  . \label{Eq:sG:static}
\end{equation}

We next assume that the phase $\mu(x,t)$ can be written in the form
\begin{equation}
  \mu(x,t)=\mu_0(x)+\sum_n\psi_n(x)e^{i\omega_n t}
  , \label{Eq:smallOsc}
\end{equation}
\ie, it performs small oscillations around the static solution $\mu_0(x)$. The eigenfunctions $\psi_n(x)$ and the eigenfrequencies $\omega_n$ are found as solutions of the Schr\"odinger equation
\begin{equation}
  -\psi_{xx} + j_c(x)\cos(\mu)\psi = \omega^2 \psi
  . \label{Eq:Schr}
\end{equation}
From all eigenfrequencies we choose the lowest one $\omega_0$ and plot it as a function of $\gamma$ in Fig.~{\ref{spectroscopy_vergl_simu}}. To plot the eigenfrequency in physical units, we have to multiply our simulation results obtained in normalized units by the plasma frequency. In Fig.~{\ref{spectroscopy_vergl_simu}} we use the scaling which provides the best fit to the experimental data. From Fig.~{\ref{spectroscopy_vergl_simu}} we easily see why our experimental data are reproduced by the simple short junction model by multiplying the obtained eigenfrequencies with appropriate integers. The experimental data are located in a parameter range where both descriptions --- the simple short junction model and the more accurate numerical simulations --- coincide. Length effects and the signature of the \ZeroPi boundary should be observable in a parameter range which is not accessible with our experimental setup. 

Thus, in this experiment we could experimentally determine the eigenfrequencies of a \ZeroPi junction and its two reference junctions. In a parameter range close to the critical current the data can be analyzed using the short junction model Eq.~(\ref{nu}).

\section{Conclusions}
In this paper, we presented the results of escape rate measurements and microwave spectroscopy of a $0$, $\pi$, and \ZeroPi ferromagnetic Josephson junction. The escape rate measurements were performed in the temperature range between 1.9\,K...20\,mK. The standard deviation $\sigma$ of the switching current distributions decreased with decreasing temperature and showed a saturation below $T^*_{{\rm exp}}\sim 150$\,mK, which is almost an order of magnitude higher than the theoretically expected temperature of the crossover to the quantum fluctuations of the Josephson phase. Thus, $T^*_{{\rm exp}}$ seems to be of a different origin. We gave arguments that the unexpected high crossover temperature is not due to setup limitations, but is an intrinsic feature of our SIFS samples. The relation between $T^*_{{\rm exp}}$ and the thermal-to-quantum transition temperature needs further investigations. Over a wide temperature range, the distribution width $\sigma^*$ of the \ZeroPi junction has larger values than the $0$ and $\pi$ junctions, possibly due to fluctuations of the fractional vortex located at the \ZeroPi boundary. 

Furthermore, we determined the eigenfrequencies of our samples experimentally by microwave spectroscopy. We observed harmonic, subharmonic and superharmonic pumping and compared our experimental data with numerical simulations of the lumped junction model.

\begin{acknowledgments}

Financial support by the Studienstiftung des Deutschen Volkes (J.~Pfeiffer) and by the DFG via SFB/TRR-21 and US\_18/10 is gratefully acknowledged. M.~Weides is supported by the project WE 4359/1-1 and the AvH foundation. 
  
\end{acknowledgments}

\bibliography{MyJJ,LJJ,SFS,SF,pi,QuComp,this2}

\end{document}